\documentclass[prl, twocolumn, showpacs, superscriptaddress]{revtex4-2}
\usepackage{amsmath, amssymb}
\usepackage{graphicx}
\renewcommand{\section}[1]{{\par\it #1.---}\ignorespaces}

\begin{document}
\title{Topological or not? A unified pattern description in the one-dimensional anisotropic quantum XY model with a transverse field}
\author{Yun-Tong Yang}
\affiliation{School of Physical Science and Technology, Lanzhou University, Lanzhou 730000, China}
\affiliation{Lanzhou Center for Theoretical Physics $\&$ Key Laboratory of Theoretical Physics of Gansu Province, Lanzhou University, Lanzhou 730000, China}
\author{Hong-Gang Luo}
\email{luohg@lzu.edu.cn}
\affiliation{School of Physical Science and Technology, Lanzhou University, Lanzhou 730000, China}
\affiliation{Lanzhou Center for Theoretical Physics $\&$ Key Laboratory of Theoretical Physics of Gansu Province, Lanzhou University, Lanzhou 730000, China}
\affiliation{Beijing Computational Science Research Center, Beijing 100084, China}

\begin{abstract}
The nature of phase transitions involving the questions why and how phase transitions take place has not been sufficiently touched in the literature.  In contrast, the current attention to certain extent still focus on the description of critical phenomena and the classification of the associated phase transition along with the Ginzburg-Landau-Wilson paradigm, where the key issue is to identify phenomenologically order parameters and related symmetries. This brings the question to topological phase transitions (TPTs), where no obvious order parameter and the broken symmetry are identified. Here we present a unified pattern description of the second-order quantum phase transition (QPT) and TPT, both involved in the one-dimensional anisotropic quantum XY model in a transverse field, which contains the transverse Ising model (TIM) as a limit case. Away from the TIM, the XY model enters the ferromagnetic phase (marked by a second-order QPT or a direct TPT) as increasing ferromagentic exchange coupling, a series of TPTs occur, which are absent in the TIM. The TPTs behave like the first-order QPTs. In the isotropic and large exchange coupling cases, the ground state of the XY model is dominated by two topologically different vortices along positive and negative direction of the transverse field. We confirm the above conclusion by analyzing the energy contributions of the patterns to the ground state and calculating the ground state pattern occupations of the XY model. The results have been obtained in a unified and self-evident way and answer the questions why and how the QPT and TPTs take place in the XY model.
\end{abstract}
\maketitle

\section{Introduction}
Topology is a mathematical concept describing originally the geometrical property of three-dimensional objects with surfaces involving possibly zero, one, or more holes. While the situation with zero hole is called topologically trivial, the rest topologically nontrivial. Irrespective of the fact that the original and intuitive picture has been greatly abstrated (e.g. the concept of quantum topology \cite{Wen2017}) in treating certain thing as a whole, the key idea of topological equivalence remains unchanged, namely, the different topologies cannot change continuously to one another \cite{Haldane2017}. 

Since 1980's, in which the finding of the integer quantum Hall effect by von Klitzing \cite{Klitzing1980} and its interpretation \cite{Thouless1982} triggered the study of quantum-topological matter, condensed matter physics enters a new stage, in which topology plays a central role in exploring the exotic properties of quantum matter or materials \cite{Hasan2010, Qi2011, Wen2017}. Here we do not plan to address such advanced issues, but would like to go a step back to wonder why and how the topology occurs in the one-dimensional (1D) quantum spin systems. Few works addressed this issue, as few works addressed the nature of  phase transitions \cite{Yang2023a, Kastner2008}. Such issue is really needed to clarify in essence. Otherwise we will encounter more and more complicated situations in understanding the physics of phase transitions including classical \cite{Landau1937, Ginzburg1950, Wilson1974}, quantum \cite{Hertz1976, Sachdev2011}, dynamical \cite{Hohenberg1977} as well as topological \cite{Berezinskii1970, Berezinskii1971, Kosterlitz1972, Kosterlitz1973} ones, just to name a few.

Focusing on the topological phase transition (TPT), known as Berezinskii-Kosterlitz-Thouless (BKT) phase transition \cite{Berezinskii1970, Berezinskii1971, Kosterlitz1972, Kosterlitz1973} is quite different from those formulated by the conventional Ginzburg-Landau-Wilson paradigm \cite{Landau1937, Ginzburg1950, Wilson1974}, also followed by quantum and dynamical phase transitions. The first and key step for this paradigm is to identify phenomenologically the order parameter and broken symmetry. As a matter of fact, this is still mainstream of current study involving the phase transition \cite{Hohenberg2015}. However, it encounters the question that no obvious order parameter and associated broken symmetry exist in the TPTs, thus the TPTs cannot fall into the conventional Ginzburg-Landau-Wilson paradigm, but are assumed to be driven by topological excitations such as the vortices with non-zero value of the vorticity in the classical 2D XY model \cite{Berezinskii1970, Berezinskii1971, Kosterlitz1972, Kosterlitz1973}. 

Turn to the quantum case, it was generally shown that due to the additional path or time history a $d$-dimensional quantum system can be equivalently treated as $(d+1)$-dimensional classical one \cite{Sondhi1997}. The key feature is the quantum fluctuations popularly existing in quantum systems due to non-commutating relations of quantum mechanical operators. Thus both the one-dimensional (1D) transverse Ising model (TIM) and the 1D quantum XY model can exhibit quantum phase transition (QPT) driven by quantum fluctuations. However, it is well known that the former is topologically trivial and the latter is topologically non-trivial in the sense that the corresponding classical 2D XY model has topological excitations \cite{Berezinskii1970, Berezinskii1971, Kosterlitz1972, Kosterlitz1973}, which has not been realized in early exact solutions of the 1D quantum XY model with antiferromagnetic exchange coupling \cite{Lieb1961} by using Jordan-Wigner transformation \cite{Jordan1928}. This is obviously not the case addressed by Haldane that the realiztion of a topological state was missed in the previous studies of quantum spin systems that ``almost all previously theoretically-described states were topologically trivial'' \cite{Haldane2017}. So the question arises what is the origin of the topology in the 1D quantum XY model and why it is absent in the 1D TIM.

Here we would like to point out that another important feature, not particularly emphasized in the literature, is the propagation of local quantum fluctuations through spatial dimension by essential interactions. This feature is absent in the TIM, in which the spin alignments influence only the energy of related spin configurations and the quantum fluctuations are only local. In contrast, in the 1D quantum XY model, the quantum fluctuations can be propagated in space due to the presence of the nearst neighbor interactions in the spin $y$-direction. As a consequence, the vortex excitations can be driven by the spatial propagation of quantum fluctuations. We employ the pattern language \cite{Yang2022c, Yang2023a, Yang2023b,Yang2023c} to explore how the topology emerges in the 1D anisotropic quantum XY model, thus indicate why and how the TPTs take place in this model.

\section{Model and Method}
The Hamiltonian of the 1D anisotropic XY model with a transverse field reads
\begin{equation}
\hat{H}' = - J'_x\sum_{i}\hat\sigma^x_i \hat\sigma^x_{i+1} - J'_y\sum_{i}\hat\sigma^y_i \hat\sigma^y_{i+1}- g\sum_i \hat\sigma^z_i, \label{XY0}
\end{equation}
where $J'_x \geq J'_y \geq 0$ represent the nearest-neighbor ferromagnetic exchange interactions along $x$- and $y$-directions of the spin located at site $i$, and $g$ the transverse field applied along the spin $z$-direction. We take $g$ as units of energy in this work and define an anisotropic parameter $\gamma = (J_x-J_y)/J$, where $J = J_x + J_y$, $J_x = J'_x/g$, $J_y = J'_y/g$, thus $\gamma \in [0, 1]$.  Eq. (\ref{XY0}) is reformulated as $\hat{H}'= \frac{g}{2} \hat{H}$, where 
\begin{eqnarray}
\hat{H} = - \sum_i \left[J (1+\gamma)\hat\sigma^x_i \hat\sigma^x_{i+1} + J(1-\gamma)\hat\sigma^y_i \hat\sigma^y_{i+1} + 2\hat\sigma^z_i\right]. \label{XY1}
\end{eqnarray}
Obviously, $\gamma = 1$ recovers the TIM and $\gamma = 0$ is the standard isotropic XY model. For the chain length $L$ with periodic boundary condition (PBC), the model Hamiltonian can be rewritten as
\begin{eqnarray}
\hat{H} &=& \left(
\begin{array}{ccccccc}
 \hat{\sigma}^x_1 &-i\hat{\sigma}^y_1&  \hat{\sigma}^x_2 & -i\hat{\sigma}^y_2& \cdots & \hat{\sigma}^x_L & -i\hat{\sigma}^y_L
\end{array}
\right)\nonumber\\
&\times&
\left(
\begin{array}{ccccccc}
0 &-1 &-J_+ &0 & \cdots &-J_+ &0 \\
-1 &0 &0 &- J_- & \cdots &0 &- J_- \\
-J_+ &0 &0 &-1 &\cdots &0 &0 \\
0 &- J_- &-1 &0 &\cdots &0 &0 \\
\vdots &\vdots &\vdots &\vdots &\ddots &\vdots &\vdots \\
-J_+ &0 &0 &0 &\cdots &0 &-1 \\
0 &- J_- &0 &0 &\cdots &-1 &0 
\end{array}
\right)\nonumber\\
&&\times \left(
\begin{array}{ccccccc}
 \hat{\sigma}^x_1& i\hat{\sigma}^y_1&  \hat{\sigma}^x_2& i\hat{\sigma}^y_2&  \cdots & \hat{\sigma}^x_L& i\hat{\sigma}^y_L
\end{array}
\right)^T,\label{XY2}
\end{eqnarray}
where $J_{\pm} = \frac{J(1\pm \gamma)}{2}$ and the identity of Pauli matrices $\hat{\sigma}^x \hat{\sigma}^y = i \hat{\sigma}^z (\text{here } i^2 = -1)$  has been used for each site and the matrix in Eq. (\ref{XY2}) has a dimension of $2L \times 2L$. It can be diagonalized to obtain eigenvalues and corresponding eigenfunctions $\{\lambda_n, u_n\} (n = 1, 2, \cdots, 2L)$, which define the patterns marked by $\lambda_n$. With these patterns at hand, Eq. (\ref{XY1}) is rewritten as
\begin{equation}
\hat{H} = \sum_{n=1}^{2L} \lambda_n \hat{A}^\dagger_n \hat{A}_n, \label{XY3a} 
\end{equation}
where each pattern $\lambda_n$ composes of single-body operators
\begin{equation}
\hat{A}_n = \sum_{i=1}^{L}\left[u_{n,2i-1} \hat{\sigma}^x_{i} + u_{n,2i} (i\hat{\sigma}^y_{i}) \right]. \label{XY3b}
\end{equation}
The validity of Eq. (\ref{XY3a}) can be confirmed by inserting into the complete basis $|\{\sigma^z_i\}\rangle (i = 1, 2,\cdots, L)$ with $\hat{\sigma}^z_i |\{\sigma^z_i\}\rangle = \pm_i(\uparrow,\downarrow) |\{\sigma^z_i\}\rangle$, as done in Refs. \cite{Yang2022c, Yang2023a, Yang2023b, Yang2023c}.

\begin{figure}[tbp]
\begin{center}
\includegraphics[width = \columnwidth]{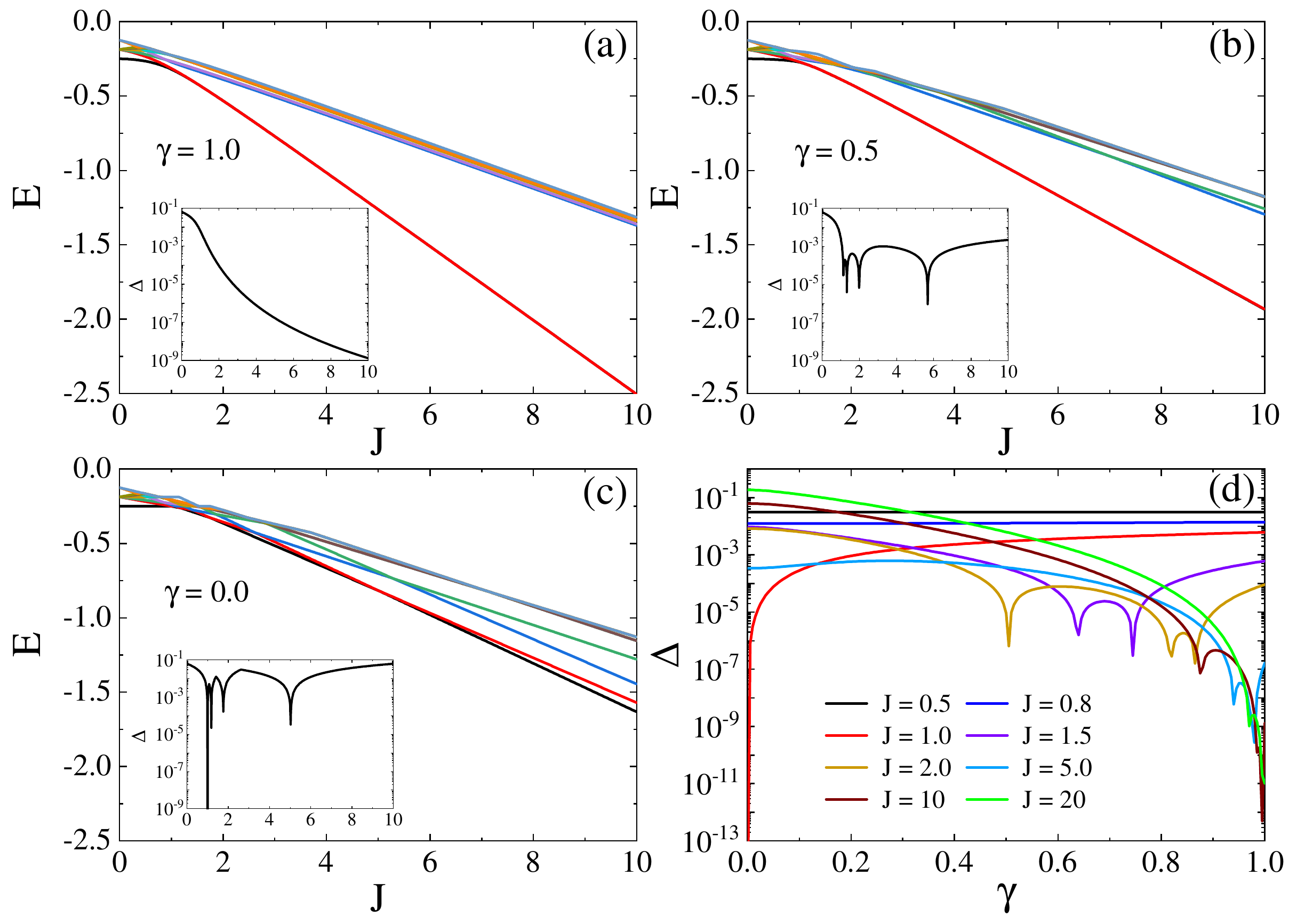}
\caption{The first eleven energy levels of the 1D anisotropic quantum XY model with $L = 8$ under PBC for different anisotropic parameter (a) $\gamma = 1$ (the TIM), (b) $0.5$, and (c) $0.0$ (the isotropic XY). The insets show the energy differences $\Delta = E_1-E_0$ between the first excited state (red solid line) and the ground state (black solid line). (d) The energy differences $\Delta$'s as functions of the anisotropic parameter $\gamma$ for different exchange coupling $J$'s.}\label{fig1}
\end{center}
\end{figure}

Figure \ref{fig1} shows the first eleven energy levels as functions of the exchange coupling $J$'s for three anisotropic parameters (a) $\gamma = 1.0$ (identical to the TIM), (b) $0.5$ and (c) $0.0$ (isotropic XY model) for $L = 8$, used in the present work. Our attention is focused on the phase transition, which can be seen from the merging behavior $(\Delta = E_1 - E_0)$ of the first excited and the ground states, as shown in the inset of Fig. \ref{fig1} (a), where a second-order QPT from the paramagnetic to ferromagnetic phases around $J \sim 1$, a well established fact \cite{Sachdev2011}. The critical point is marked by the dip of the second derivative of the ground state energy, as shown in Fig. \ref{fig4} (a2) later. Away from the TIM, besides the second-order QPT, four dips occur in the ferromagnetic phase of the system, which is obviously absent in the TIM, as observed from the behavior of gap $\Delta = E_1 - E_0$ as a function of $J$ in the insets of \ref{fig1} (b), (c). Moreover, these dips move toward the direction of smaller $J$'s, and finally, one of them approaches to $J=1$, thus inundates the second-order QPT, as shown in the insets of Fig. \ref{fig1} (b) and (c), which have not been reported in the literature. These dips demonstrate the closes or closing trends of the energy gap $\Delta$'s, in sharp contrast to the simple merging behavior in the case of the TIM ($\gamma = 1.0$). More richer dip structures are presented in Fig. \ref{fig1} (d) for different $J$'s, it begins from $J = 1.0$ at $\gamma = 0.0$ and almost disappears for enough large $J$ such as $J = 20$ around $\gamma \sim 1.0$. The closes or closing trends of the energy gap $\Delta$'s are closely related to the behavior of the pattern $\lambda_9$ and the topological phase transition, as discussed below. 

\begin{figure}[tbp]
\begin{center}
\includegraphics[width = \columnwidth]{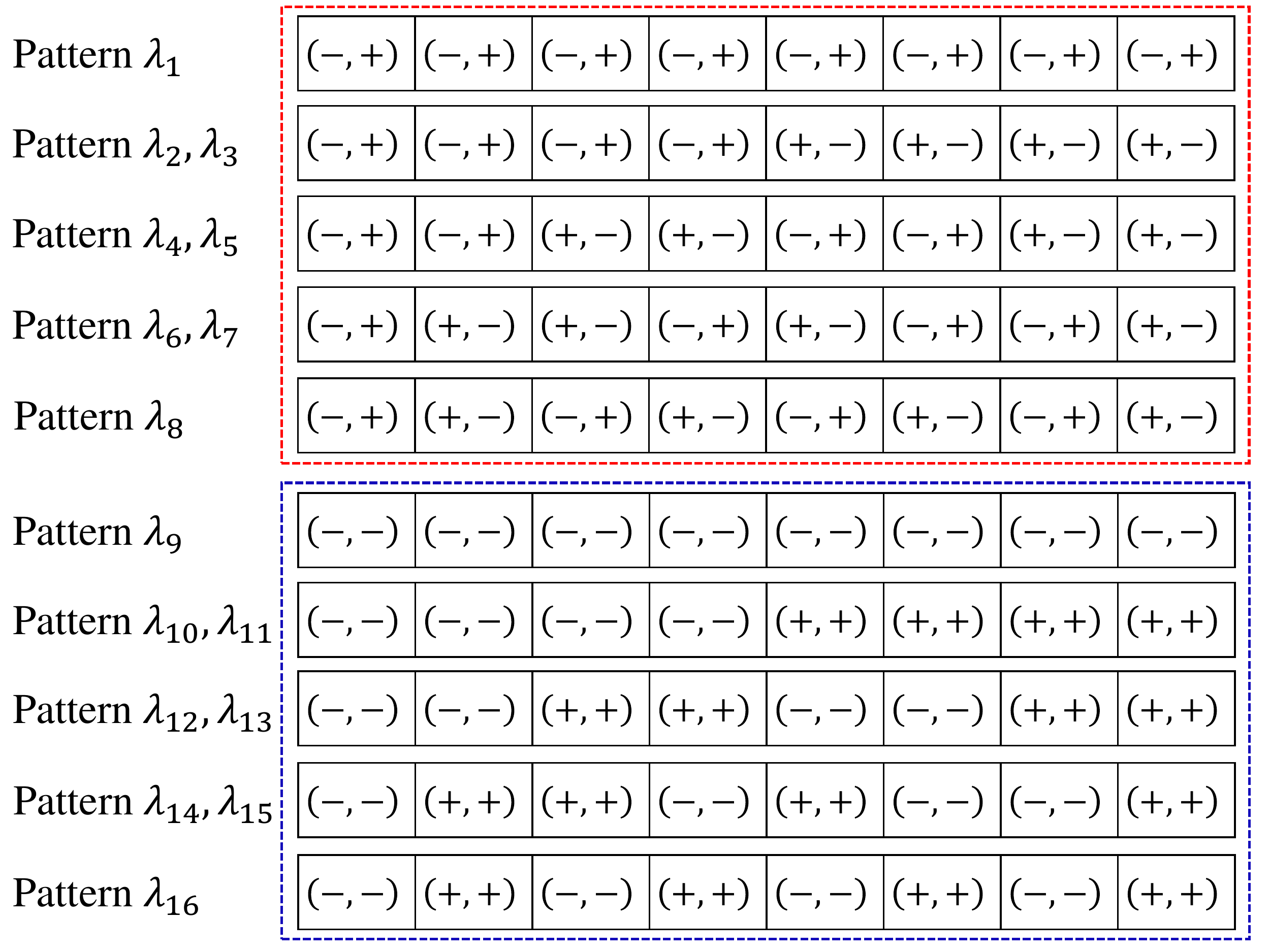}
\caption{The patterns marked by $\lambda_n (n = 1, 2, \cdots, 2L)$ consisting of the signs of the coefficients of $\hat{\sigma}^x$ and $i\hat{\sigma}^y$ for each site, which are given by Eq. (\ref{XY3b}). The values of the coefficients themselves (not shown here) are just monotonic or constant functions of the exchange coupling $J$.}\label{fig2}
\end{center}
\end{figure}

\begin{figure}[tbp]
\begin{center}
\includegraphics[width = \columnwidth]{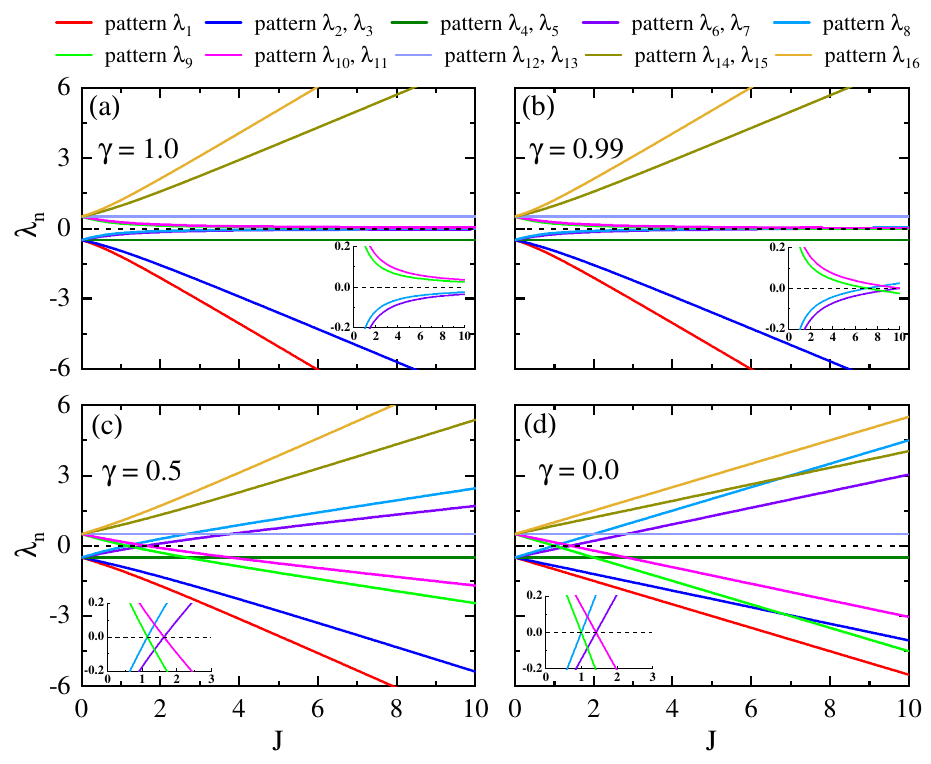}
\caption{The eigenenergies of the patterns as functions of the exchange coupling $J$ for different anisotropic parameter (a) $\gamma = 1.0$ (the TIM), (b) $0.99$, (c) $0.5$, and (d) $0.0$ (the isotropic XY). The insets show the enlarged views near zero energy.}\label{fig3}
\end{center}
\end{figure}

\begin{figure}[tbp]
\begin{center}
\includegraphics[width = \columnwidth]{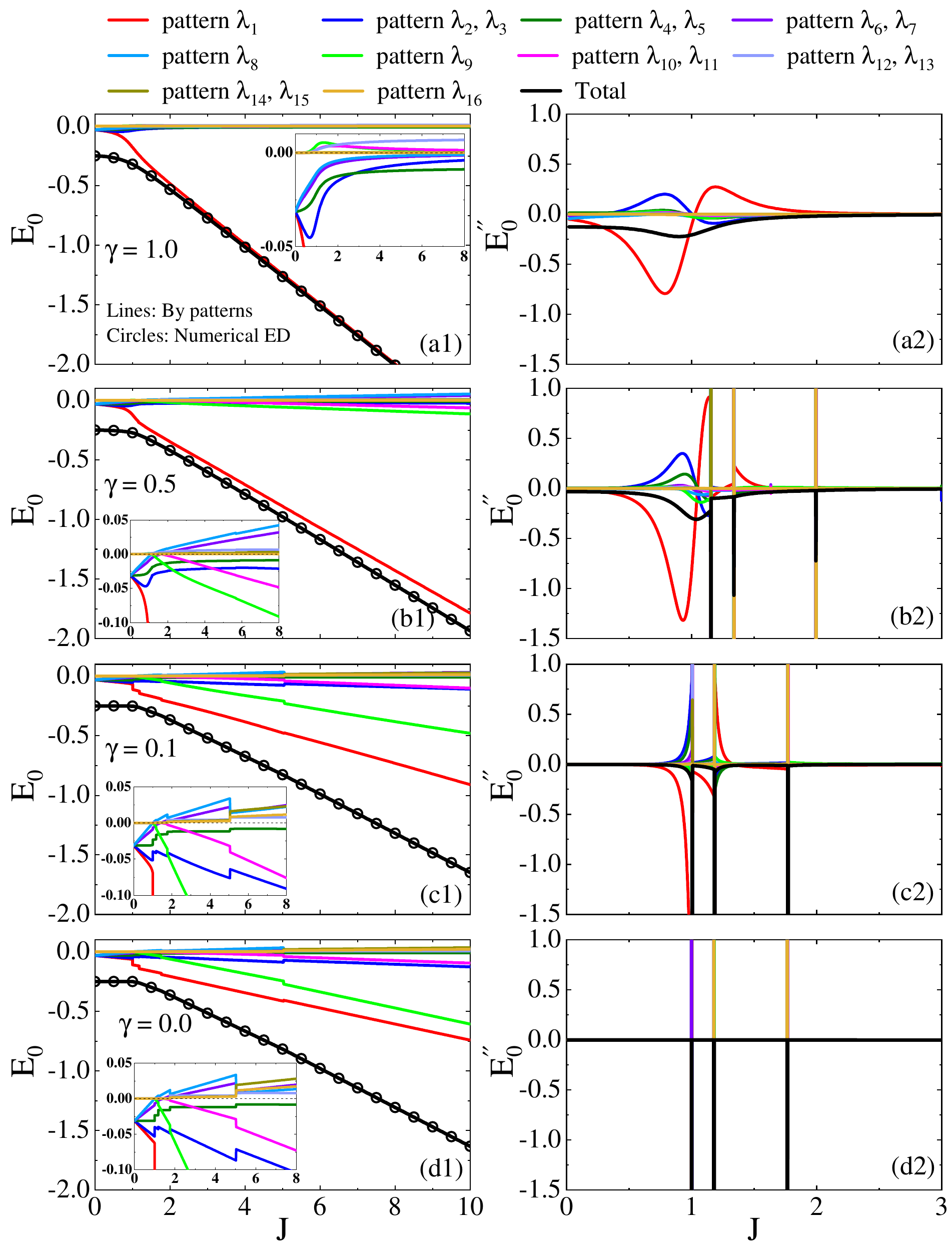}
\caption{(a1)-(d1) The ground state energies (thick black solid lines) and the contributions of patterns (thin colored solid lines) as functions of the exchange coupling $J$; (a2)-(d2) the second derivatives of the ground state energies (thick black solid lines) and the corresponding pattern components (thin colored solid lines). (a1) $\&$ (a2) $\gamma = 1.0$ (the TIM); (b1) $\&$ (b2) $\gamma = 0.5$; (c1) $\&$ (c2) $\gamma = 0.1$; (d1) $\&$ (d2) $\gamma = 0.0$ (the isotropic XY). The insets in (a1)-(d1) show the enlarged view of the contributions of various patterns.}\label{fig4}
\end{center}
\end{figure}

\begin{figure}[tbp]
\begin{center}
\includegraphics[width = \columnwidth]{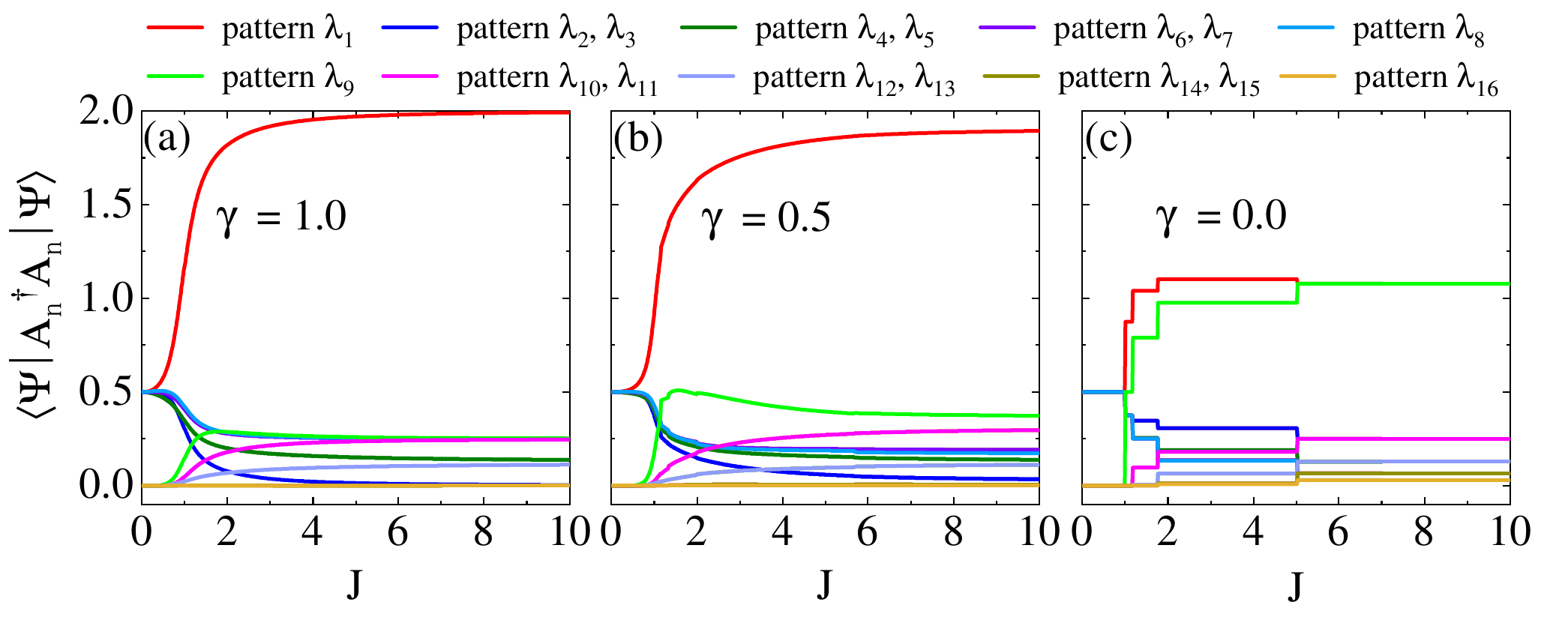}
\caption{The patterns' occupancy in the ground state of the system for different anisotropic  parameter $\gamma$'s as functions of the exchange coupling $J$. }\label{fig5}
\end{center}
\end{figure}

\section{Topological phase transitions} 
The marks of the patterns are shown in Fig. \ref{fig2}, and they are the signs of the coefficients of $\{\hat{\sigma}^x_i, i\hat{\sigma}^y_i\}$ in Eq. (\ref{XY3b}). They denote the in-phase if the signs are same and out-of-phase if not, having same physical meanings as those of Fig. 1 in Ref. \cite{Yang2023a}. The eigenenergies of the patterns are shown in Fig. \ref{fig3} here. An obvious feature in the TIM ($\gamma = 1.0$) is that there are no level crossings, as shown in the inset of Fig. \ref{fig3} (a). Once the anisotropic $\gamma$ is lesser than $1.0$, the $J_y$ component comes into play, as a consequence, the level crossings occur, as shown in the inset of Fig. \ref{fig3} (b) and the crossing points move towards the smaller $J$ with decreasing $\gamma$, up to $J = 1.0$ for $\gamma = 0.0$. These observations are clear from Fig. \ref{fig3} (c) and (d) and the insets in them. The eigenenergies crossings, specially those of the pattern $\lambda_9$ play an important role in the TPTs and we discuss it in detail in the following.

Figure \ref{fig4} shows the ground energies [thick black solid lines in (a1)-(d1)] of the system for several typical anisotropic parameter $\gamma$'s and corresponding pattern contributions (thin colored solid lines). In order to confirm our results, numerical exact diagonalization (ED) results are also presented (circles). The exact agreements have been obtained. This is not surprising since no any approximation has been introduced for the present calculations. The insets show enlarged view of the contributions for various patterns. It is noted that (i) for large $\gamma$ the contribution from the pattern $\lambda_1$ dominates over the others; (ii) for small $\gamma$, specially $\gamma = 0.0$ the contribution from the pattern $\lambda_9$ becomes comparable to that of the pattern $\lambda_1$; (iii) at the same time, the contributions from the other patterns become sizeable, in particular with increasing the exchange coupling $J$'s; (iv) the ground state of the system shows a second-order QPT, occurring at $J \sim 1$ for $\gamma = 1.0$ (the TIM); (v) this second-order QPT remains visible up to $\gamma = 0.5$, but it is inundated by approaching first-order-like QPT with further decreasing $\gamma$, for example, in the case of $\gamma = 0.1$, this second-order QPT becomes unvisible and (vi) it disappears completely in the case of $\gamma = 0.0$, in which the remainings are four sharp transitions marked by the singularities of the second derivatives shown in Fig. \ref{fig4} (b2)-(d2). These are TPTs, which are discussed in details below. 

It is clear that the second-order QPT denotes that the system enters the ferromagnetic phase, as well-known in the TIM ($\gamma = 1.0$) \cite{Sachdev2011}. The same thing also happens for $\gamma < 1.0$, but the difference to that of $\gamma = 1.0$ is that the role of the pattern $\lambda_9$ comes into play and becomes more and more important, in particular for $\gamma = 0.0$. From Fig. 2 the pattern $\lambda_1$ is marked as the signs of $\{\hat{\sigma}^x_i, i\hat{\sigma}^y_i\}: \{(-,+),(-,+), \cdots, (-,+)\}$ and the pattern $\lambda_9$ marked as $ \{(-,-),(-,-), \cdots, (-,-)\}$, both which exist in the TIM ($\gamma = 1.0$ here) and in the isotropic XY model ($\gamma = 0.0$). Obviously, from the $\hat{\sigma}^x$ component the pattern $\lambda_1$ and $\lambda_9$ at all denote the ferromagnetic phase, namely, all spins in different sites are in-phase. The difference is the phase of the spin components at each site: they are out-of-phase in the pattern $\lambda_1$, but in-phase in the pattern $\lambda_9$. This represents that the states denoted by the patterns $\lambda_1$ and $\lambda_9$ are different \textit{topologically}. While the pattern $\lambda_9$ does not come into play as an important role in the TIM ($\gamma = 1.0$), it plays a comparable role in the isotropic XY model in comparison to the pattern $\lambda_1$. In this sense the TIM is not topological, but the isotropic XY model is, additionally taking into account the 2D nature of the 1D quantum model \cite{Sondhi1997}: comparable vortex and anti-vortex cannot transform to one another in a continuous way. This topology leads to unusual feature in the ferromagnetic phase of the XY model: a series of TPTs take place once the pattern $\lambda_9$ goes across the other patterns, as shown as singular points in the second derivatives of the total ground state energies and their pattern components, which are obviously absent in the TIM ($\gamma = 1.0$). These singular points are in accordance with the eigenenergies crossing positions of the pattern $\lambda_9$ with the others (there are also other eigenenergies crossings but they are unimportant due to their smaller contributions in comparison to that of the pattern $\lambda_9$ in the case of $\gamma < 1.0$). These singular behaviors are completely consistent with the dips (energy gap closings) observed in the insets of Fig. \ref{fig1}, which is characteristic behavior of TPTs.

Figure \ref{fig5} presents a comparison of the pattern occupations in the ground state of the models in different phases determined by the ferromagnetic exchange coupling $J$'s. The result for the TIM ($\gamma = 1.0$) is consistent with that shown as histogram in Ref. \cite{Yang2023a}. Once the second-order QPT occurs at $J = 1$, the occupation of the pattern $\lambda_1$ becomes dominated, and the rest has less contributions, including that of the pattern $\lambda_9$, where no any TPT occurs. With decreasing $\gamma$, the contribution from the pattern $\lambda_1$ is suppressed in a significant way and at the same time that of the patterns $\lambda_9$ becomes more and more apparent. In particular, at $\gamma = 0.0$, it becomes comparable to that of the pattern $\lambda_1$. At the same time, the occupations of various patterns have several dramatic steps, which are in consistent with the TPTs, occurring once the eigenenergy of the pattern $\lambda_9$ crosses the other patterns. This is also the reason why the TPTs look more like the first-order QPT. As mentioned above, these two patterns are topologically different in nature, which indicate why and how the TPTs take place, and thus reach the topological ground state of the system containing mainly two competitive ferromagnetic topological orders.

\section{Summary and Discussion}
We have investigated the second-order QPT and possible TPTs in the 1D anisotropic quantum XY model, in comparison to the case of the TIM where the TPT is absent. The pattern language provides a natural and powerful way to describe these phases and phase transitions in a self-evident way in the sense that the patterns denote directly the phases and the crossings of the pattern eigenenergies denote directly the possible TPTs, associated with closes or closing trends of energy gaps, an essential feature of TPTs. 

The above discussion is made for finite lattice size ($L = 8$ here) and extensions toward more larger sizes or higher dimensions are straightforward, and the essential physics discussed here remains valid but more computational cost is needed. However, one can obtain some intuition only from the patterns obtained by diagonalizing the Hamiltonian matrices, just like Eq. (\ref{XY2}). More physics on the topological nature of the XY model is also interesting, which is left for future study.

\section{Acknowledgments}
The work is partly supported by the National Key Research and Development Program of China (Grant No. 2022YFA1402704) and the programs for NSFC of China (Grant No. 11834005, Grant No. 12247101).



%

\end{document}